# Physics of band-filling correction in defect calculations of solid-state materials

Harshan Reddy Gopidi[1], Lovelesh Vashist[1], and Oleksandr I. Malyi[1,#]

[1]Centre of Excellence ENSEMBLE3 Sp. z o. o., Wolczynska Str. 133, 01-919, Warsaw, Poland

[#]Email: oleksandrmalyi@gmail.com

**Abstract:**
In solid-state physics/chemistry, a precise understanding of defect formation and its impact on the electronic properties of wide-bandgap insulators is a cornerstone of modern semiconductor technology. However, complexities arise in the electronic structure theory of defect formation when the latter triggers partial occupation of the conduction/valence band, necessitating accurate post-process correction to the energy calculations. Herein, we dissect these complexities, focusing specifically on the post-process band-filling corrections, a crucial element that often demands thorough treatment in defect formation studies. We recognize the importance of these corrections in maintaining the accuracy of electronic properties predictions in wide-bandgap insulators and their role in reinforcing the importance of a reliable common reference state for defect formation energy calculations. We explored solutions such as aligning deep states and electrostatic potentials, both of which have been used in previous works, showing the effect of band alignment on defect formation energy. Our findings demonstrate that the impact of defect formation on electronic structure (even deep states) can be significantly dependent on the supercell size. We also show that within band-filling calculations, one needs to account for the possible change of electronic structure induced by defect formation, which requires sufficient convergence of electronic structure with supercell size. Thus, this work emphasizes the critical steps to predict defect formation energy better and paves the way for future research to overcome these challenges and advance the field with more efficient and reliable predictive models.

Point defects are present in all materials as their formation arises from a balance between the energy cost needed for their formation and configuration entropy gain due to increased defect concentration.[1, 2] Despite being present only in low concentrations under normal conditions in most materials, defects often significantly influence the properties of materials, such as color, equilibrium Fermi level, and doping response.[3-5] With recent advancements in electronic structure theory, defect physics and hence material properties can be accurately predicted using first-principles calculations. However, it is important to note that density functional theory (DFT) is still limited to supercell calculations of only a few hundred atoms. Hence, when computing defect properties for typical supercells, the first-principles results should be extrapolated to the dilute limit. This can be achieved by either (i) scaling corresponding properties (e.g., vacancy formation energy) for different defects as a function of supercell size[6-9] or (ii) applying post-process corrections[10-14]. While the former approach is theoretically exact and only requires sufficient supercomputer resources, it can be difficult to apply for many defects due to the need to scale each defect individually. In contrast, the latter approach typically involves applying post-process corrections to extrapolate the limit of defect formation energy and is becoming increasingly popular in modern research.

Previous works[10-15] have shed some light on post-process corrections, and several post-process correction codes have been developed to automate defect calculations. Thus, it becomes clear that the formation energy of a point defect in an insulator is directly affected by the formation energies of other charged defects, which are a function of the parametric Fermi level. Hence, the calculation of equilibrium defect concentration or equilibrium Fermi level corresponding to given environmental conditions requires a self-consistent solution that follows the charge neutrality[3, 4, 16, 17] rule and accounts for the effect of the defect on the average electrostatic potential as well as the accounting for energy correction due to periodic interaction between point defects. Moreover, dielectric screening, which can vary anisotropically along different directions, must be considered when dealing with charged point defects.

There is a common misconception that post-process corrections are only necessary when dealing with charged defects and that uncharged defects do not require such corrections. This is only partially accurate. For instance, both charged and uncharged point defects can cause partial occupation of the conduction or valence band. Within electronic structure theory for a finite supercell size, such electrons or holes are spread in energy range. In contrast, in the dilute limit (one which we often aim to represent within electronic structure theory), they are expected to only contribute to the tip of band edges (see Fig. 1a). This thus implies that formation energy of point defect adding free carriers to the system requires accounting for post-process energy corrections accounting for additional energy cost corresponding to band-filling. This additional correction for the doped insulator can be calculated by summing over the eigenvalues[11, 13, 18] as:

For n-type dopants:
$$\Delta_{BFC} = -\Sigma_{n,k}[\Theta(e_{n,k} - E^*_{cbm})\omega_k \gamma_{n,k}(e_{n,k} - E^*_{cbm})] \quad (1)$$

For p-type dopants:
$$\Delta_{BFC} = -\Sigma_{n,k}[\Theta(E^*_{vbm} - e_{n,k})\omega_k (1 - \gamma_{n,k})(e_{n,k} - E^*_{vbm})] \quad (2)$$

where $\Theta(x)$ is the Heaviside step function, $\omega_k$ are the weights of the k-points, $e_{n,k}$ are the energy eigenvalues of state (n, k), $\gamma_{n,k}$ are the occupations of the eigenstate (n, k), $E^*_{cbm}$ is the aligned conduction band minimum (CBM) of supercell without defects, $E^*_{vbm}$ is the aligned valence band maximum (VBM) of the supercell without defects. Fig. 1b shows the expected band-filling correction due to the rigid shift of the Fermi level in the conduction band for some well-known insulators. One of the key conclusions here is that an increase in the band dispersion (often simply characterized by low effective mass) increases the expected band-filling correction. As an illustration, for effective lattice parameter (i.e., defined as cubic root of the system volume) of around 10 Å, the magnitude of band-filling correction for the system containing 1e in the conduction band is more than 0.8 eV for CdTe, ZnO, and GaAs. At the same time, the corresponding value for BN is less than 0.1 eV.

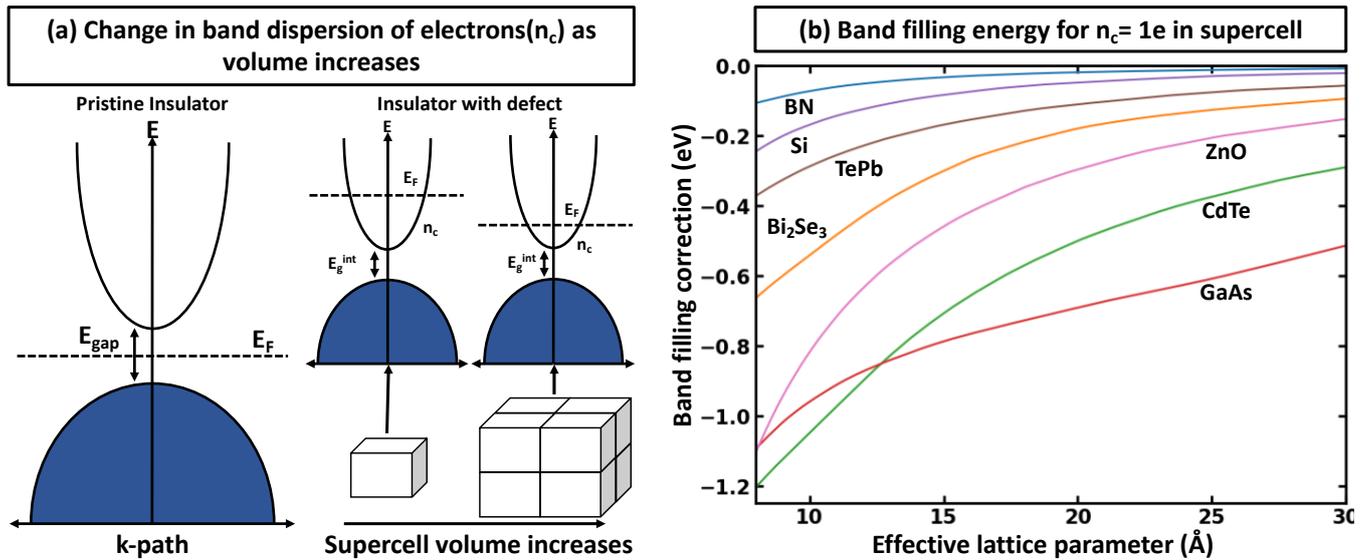

**Figure 1. Defect formation resulting in the band filling dependent on supercell size.**
(a) Origin of band-filling correction (as a consequence of the finite size of the system). (b) Band filling correction to supercells for different materials as the function effective lattice parameter (i.e., defined as the cubic root of the system volume), calculated with the rigid shift of the Fermi level in the conduction band for some well-known insulators.

We note, however, that for the defective system, Eqs. 1-2 cannot be directly used in modern first principles codes (e.g., VASP [19-22] and Quantum Expresso [23-25]) because main plane-wave DFT codes do not use common reference energy but provide direct Kohn-Sham-eigenvalues for set of k-points. To illustrate this behavior, we examine the case of defect formation in a conventional insulator. Our findings suggest that defect formation results in the shift of the absolute energy of different eigenvalues. Consequently, a direct comparison of eigenvalues for two separate calculations - such as pristine and defective systems - becomes meaningless. This suggests that the equation mentioned above necessitates the use of a common reference state. Primarily, the rationale for this correction is that the bands of defective and pristine systems must be accurately aligned. However, the formation of a point defect often results in a localized structural perturbation and could even alter the internal band gap, making it impossible to directly align the band edges. Moreover, the band energies in pristine and defective systems are not directly comparable. For example, Figure 2a demonstrates the electronic density of states for pristine and defective ZnO systems, revealing a clear shift in eigenvalue energies between the two systems. Hence, it can be inferred that comparing the electronic structure for different systems is only meaningful when a common reference state is defined. The sensitivity of defect formation energy to the way this reference state is defined is shown in Figure 2b. The question then arises as to how we define the common reference state. The answer is not straightforward and is contingent upon the specifics of the system under examination, which we will discuss further.

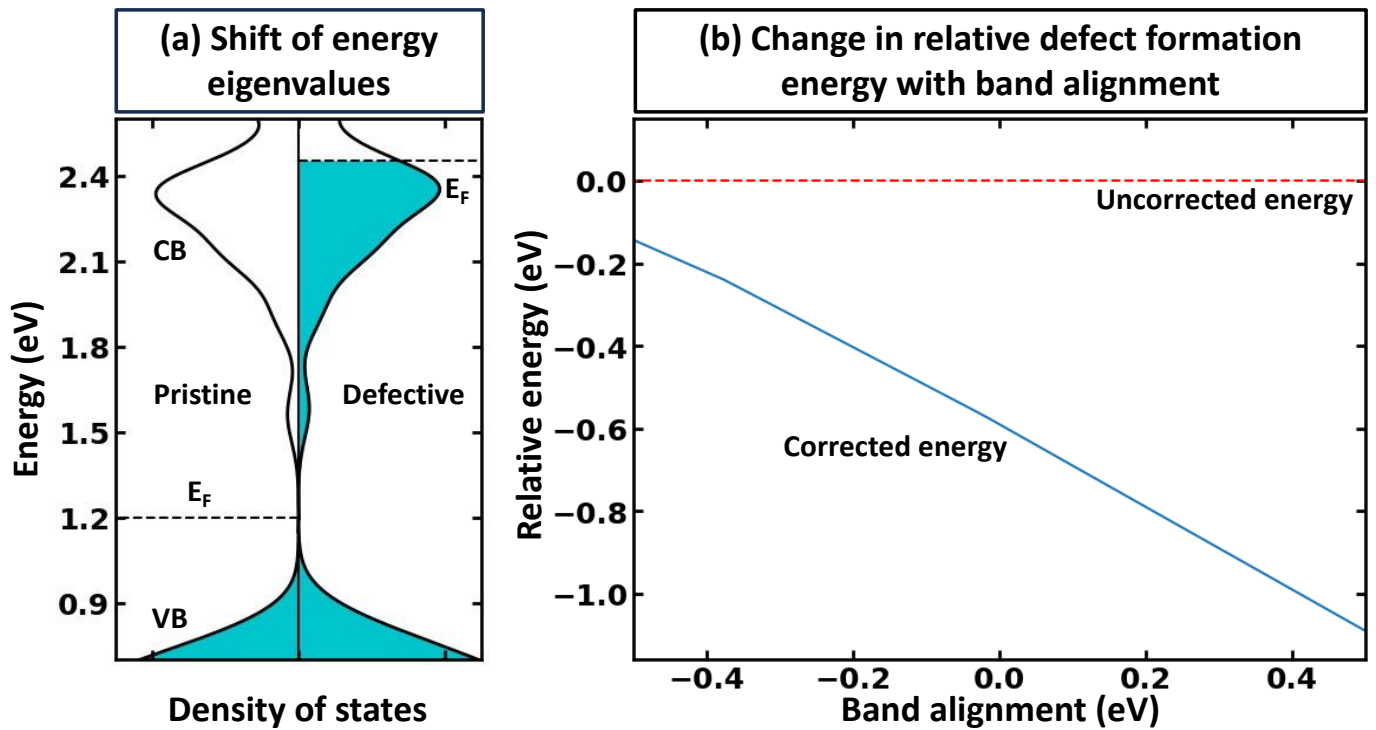

**Figure 2. Effect of defect formation on the electronic structure of ZnO.**
(a) Electronic structure of pristine ZnO and defective ZnO:Al$_{Zn}$. The defect formation results in a slight upward shift of energy eigenvalues. States that are occupied are represented with shading, while unoccupied states are indicated with white. The Fermi level for the pristine system is shown in the middle of the band gap. For visualization pursues, the density of states for principal conduction band is multiplied by 10. (b) Relative defect formation energy for Al substitutional defect (ZnO:Al$_{Zn}$) as a function of band alignment used for calculations of band filling correction. The results are shown for PBE calculations, including band-filling correction. The results are presented for a 192-atom supercell.

**Alignment of deep states (average) as the reference state:** Within the global scientific community, there exists a consensus that recognizes the significant role that the reorganization of valence states plays in the formation and nature of chemical bonds. The foundation of this principle is based on the idea that transitions in the electron states of an atom's outermost shell do not simply dictate the variety of bonds that the atom is capable of forming. In addition, these shifts also wield a considerable influence on key factors such as the strength, stability, and overall reactivity of the bonds themselves. Following this line of thought, core electrons or, more generally, deep states, which occupy lower energy levels and are somewhat insulated by the outer electron shell from direct influences of other atoms (they till indirectly influence bonding) can be used as potential consistent reference points across various systems. In the case of full electron codes (e.g., Wien2k [26]), one can explicitly calculate the position of the deepest orbitals and align them to define the common reference state. We note, however, that the most widely used first principles codes use the pseudopotentials to simplify the computational process and reduce the computational cost associated with solving the Kohn-Sham equation [27]. As a result, the position of the deep state is only 10-25 eV below the Fermi level, depending on the considered valence configuration. In Figs. 3a,b, we summarize the effect of defect formation on the deep states for a few representative compounds and supercells without any relaxation for the defective system. These results show that even deep states can have substantial broadening. For instance, for CdTe:In$_{Cd}$ described within a 54-atom supercell, the broadening is 31 meV (Fig. 3c). It should also be noted that defect formation can, in principle, cause interference of defect deep states with the states of the host atoms as seen in Fig. 3b. One might initially think that a small change or broadening of the deep states would not significantly affect the defect formation energy. However, as shown in Fig. 2b, the energy alignment does directly affect the results, and the effect is expected to increase with higher free carrier concentration introduced by the defects. This means that accurate defect calculations will require not only accounting of band-filling correction but also an accurate alignment of energy states of pristine and defective systems.

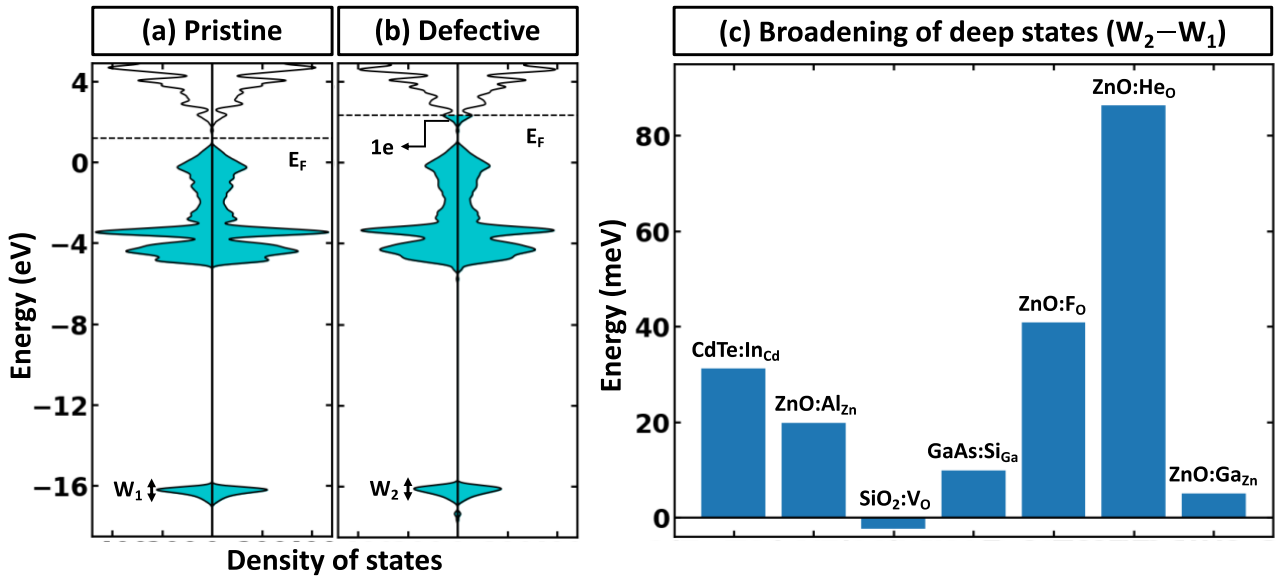

**Figure 3. Effect of defect formation on the deep states.**
Density of states of (a) pristine and (b) defective ZnO:$Al_{Zn}$ systems. The results are shown for a 192-atom supercell. For visualization pursues, the density of states for the principal conduction band is multiplied by 30. The Fermi level for the pristine system is shown in the middle of the band gap. (c) Histogram for broadening of deep states (defined here as deeper than 10 eV from the principal valence band in the pristine system) upon defect formation. The results are presented for 54, 192, 162, 54, 96, 96, and 96-atom supercells for CdTe:$In_{Cd}$, ZnO:$Al_{Zn}$, SiO$_2$:$V_O$, GaAs:$Si_{Ga}$, ZnO:$F_O$, ZnO:$He_O$, and ZnO:$Ga_{Zn}$, respectively.

**Alignment of electrostatic potential as reference state:** While defects can significantly influence the electronic properties of materials — given that these properties are highly sensitive to specific atomic arrangements, bonding situations, and shifts in the local environment[28, 29] — the electrostatic potentials within the system remain comparatively undisturbed. This stability stems from the fact that electrostatic potential is a long-range attribute derived from charge interactions spanning vast distances. Consequently, the effects of large-scale environmental shifts are averaged out, while the impact of the local environment remains crucial. As one moves sufficiently far from the defect, the local environments mirror that of the pristine system. In such regions, comparing the electrostatic potentials offers insight into energy alignment. Consequently, it is viable to use the mean electrostatic potentials — particularly, potentials that are spherically averaged around atoms within a specified distance — for two systems as a common method for energy alignment. This strategy is particularly applicable when calculating defect formation energy. We note, however, the introduction of the defect inside the supercell results in the symmetry breaking, making different atoms inequivalent to each other and resulting in the space and atomic identity-dependent electrostatic potentials. This is shown in Fig. 4a-c, where one can see that the change in relative electrostatic potential slowly converges with moving away from the defect, reaching the constant value (within 0.01 eV) at about 15 Å away from the defect, as shown by an example of ZnO:$He_O$ system (herein, modeled as 640-atom supercell)- a type of functional defect resulting in putting Fermi level to the conduction band with 2e spread over a wide energy range in the conduction band.[30, 31] We also note that the tolerance value for band alignment needed for accurate calculation of defect formation energy differs for different systems as the band filing correction is affected by the band alignment and free carrier concentration. In other words, the large free carrier concentration (intrinsic or doping induced) requires more accurate band alignment calculations. While the results are demonstrated on example ZnO, as noted in Figure 1, the effect is expected to be present for all compounds. To demonstrate the power of band-filling correction, we consider ZnO:Al$_{Zn}$, a system where the formation of the point defect (Al on Zn site) puts 1e in the conduction band. Indeed, Al is one of the standard n-type dopants in ZnO to realize n-type transparent conductive oxides.[32, 33] The uncorrected PBE defect formation energy has a strong supercell size dependence of defect formation energy (Fig. 4d), with relative defect formation energy changing by 0.5 eV as supercell size increases from 72 to 980 atoms. Importantly, when band-filling correction is accounted for (accounting for energy alignment using an electrostatic potential for atoms most remote from the defect), the defect formation energy becomes almost supercell independent.

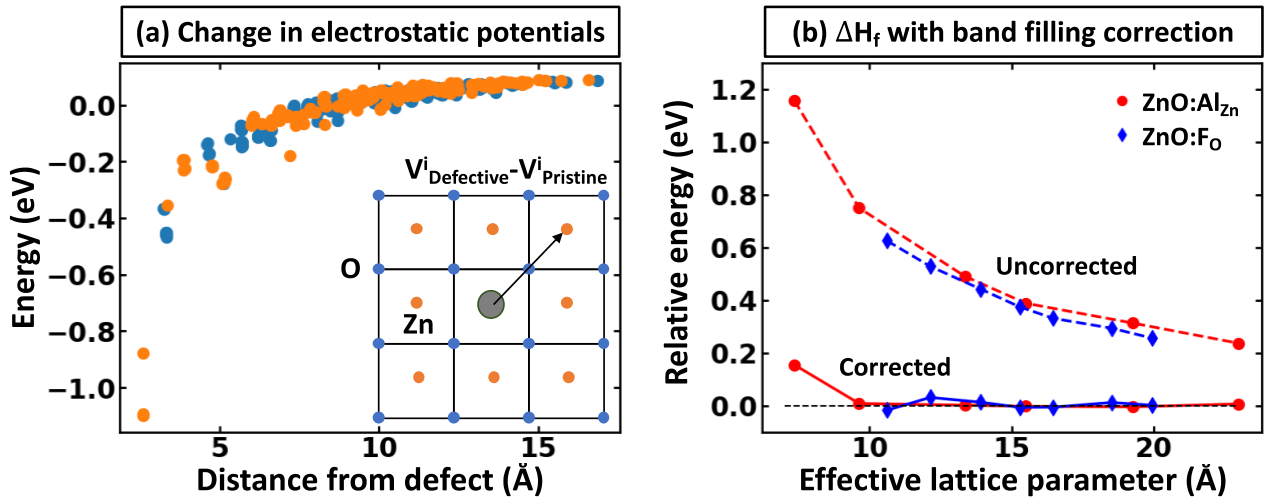

**Figure 4. Electrostatic potential as reference state for band alignment.**
(a) Change in electrostatic potentials ($V^i_{Defective}-V^i_{Pristine}$, where $V^i_{Defective}$ and $V^i_{Pristine}$ are the electrostatic potentials in defective and pristine systems, respectively) as a function of the distance (in defective supercell) from the defect shown in the example of ZnO:He$_O$ using a 640-atom supercell. (b) supercell size dependence of relative defect formation energy for ZnO:Al$_{Zn}$ and ZnO:F$_O$. For ZnO:F$_O$ and ZnO:Ga$_{Zn}$ (not shown), the results are visually not distinguishable. The results are shown for PBE calculations, including band-filling correction using energy alignment from most remote Zn/O atoms in the system (the calculation is based on the average fluctuation in electrostatic potentials for atoms positioned within spherical shells defined by two radii, $r_{max}$ and $r_{max}-r_{tol}$, here, $r_{max}$ represents the maximum radial distance of any atom within the supercell from the defect, while $r_{tol}$ is a tolerance radius set to be 2 Å).

**Other factors that should be accounted for in the post-process correction:** We must recognize that the lack of dependence of defect formation energy on the supercell size, even after considering the post-process band-filling correction, does not necessarily stand as a complete representation of accurately depicted primary physics. The eqs. 1-2 take into account the effect of band-filling on the kinetic energy of a system composed of non-interacting particles within the context of Kohn-Sham DFT. However, this method does not account for the relaxation or displacement convergence relative to the supercell size. For certain systems, this particular aspect may be profoundly significant. This is well illustrated in the scenario of noble gas functional defects that exhibit atypical relaxation patterns in solids. For instance, achieving convergence of atomic displacements in noble gas-doped ZnO often necessitates the use of a supercell comprised of more than 500 atoms.[31] Moreover, it is crucial to acknowledge that the correction detailed in Eqs. 1-2 only extends to the band edge of the pristine system. Therefore, if the defect formation results in alterations of the internal band gap, the band-filling correction will not account for these changes. Let us understand the latter case in detail. When a defect is introduced, it breaks the internal symmetry. This symmetry breaking may modify the local density of states for nearest atoms significantly, resulting in change the internal band gap energy. For instance, Fig. 5a demonstrates the convergence of the PBE internal band gap between principal band edges of ZnO:Al$_{Zn}$ as a function of supercell size. Here, one can see two main points: (i) there is supercell size dependence of internal band gap; (ii) defected system has an internal gap between principal band edges smaller than the undoped system in the dilute limit (infinite supercell). It is critical to note that the reduction of band gap energy in the dilute limit can be fully physical phenomenon originating from the formation of different motifs contributing differently to electronic properties (Fig. 5b,c). This indeed is a common phenomenon in the case of symmetry breaking in quantum materials [28, 29]. However, the change in electronic properties can also be an example of artificial interaction between the defect-induced local environment and its periodic images. Indeed, as shown in Fig. 5a, at the small size of supercells, the internal band gap energy can be substantially smaller than the asymptotic band gap energy, and then the corresponding changes require additional correction different from the standard band filling correction.

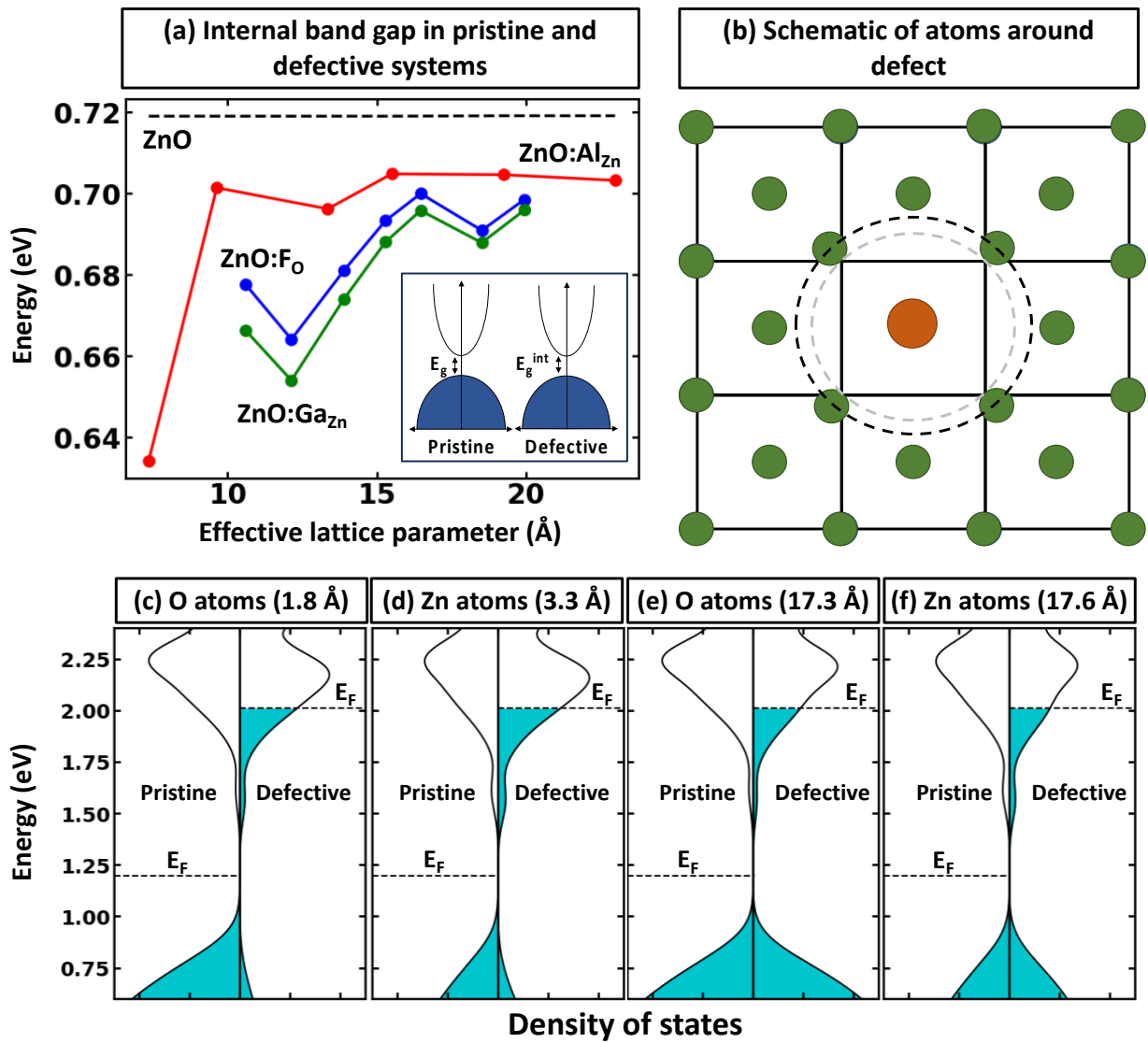

**Figure 5. Other factors that should be accounted for the post-process correction.**
(a) Illustrates the variation in the band gap as a function of supercell size, demonstrating the material response to supercell size. The results are shown for ZnO:Al$_{Zn}$, ZnO:F$_O$, ZnO:Ga$_{Zn}$ and ZnO supercells. (b) Shows schematically the local perturbations of atoms near the defect. Panels (c) to (f) represent the projected density of states, after band alignment, at different distances form the defect position in ZnO (on the left) and ZnO:Al$_{Zn}$ (on the right) supercells comprising 980 atoms. These panels provide insights into the electronic state distribution in various atomic spheres of the system. The Fermi level for the pristine system is shown in the middle of the band gap. For visualization pursues, the projected density of states for the principial conduction band are multiplied by 40.

In summary, this study highlights the necessity of applying post-process band-filling corrections in defect energy calculation when defects trigger partial occupation of the conduction or valence band. We underline the need for a robust, common reference state (energy alignment) in defect formation energy calculations, which becomes challenging given the local perturbation of structure and potential changes to the internal band gap upon defect formation. We have explored possible solutions, including aligning deep states and electrostatic potentials (which are used in some of the previous works), but have also highlighted the inherent difficulties in each approach. We have also pointed out that the impact of defect formation on electronic properties can be notably supercell size-dependent in some cases. This study thus serves as an important stepping stone towards a more comprehensive understanding of defect formation and electronic structure theory. It underscores the urgency of developing more accurate, encompassing methodologies to better predict and account for these dynamics in the context of varying supercell sizes and defect-induced alterations. Future research should tackle these challenges to advance the field and realize more efficient and reliable predictive models.

**Methods:** All calculations at the first-principle level were performed using the Vienna Ab initio Simulation Package (VASP)[19-22], employing the Perdew-Burke-Ernzerhof (PBE)[34] functional. For plane wave basis, the cutoff energy levels were set to 520 eV and 500 eV for volume relaxation and structural relaxation, respectively (except for $ZnO:F_O$, $ZnO:He_O$ and, $ZnO:Ga_{Zn}$ calculations of volume and structural relaxations, if performed, are done with 550 and 450 eV respectively). The atomic relaxations were carried out (unless otherwise specified) until the intrinsic forces were below 0.01 eV/Å. The Γ-centered Monkhorst–Pack k-grid [35] with density of 10,000 per reciprocal atom was used for all main calculations. The results were analyzed using Pymatgen[36] and Vesta[37]. It should be noted that PBE exchange-correlation functional tends to underestimate the band gap energy compared to experimental data.[38, 39] This discrepancy is fundamentally due to the limitations of generalized gradient approximations. This underestimation can directly affect defect properties.[40, 41] However, the focus of this work is not to attain precise calculations of defect formation energies, but rather to develop a deeper understanding of the band-filling correction. Even with the band gap underestimation issue inherent to the PBE functional, it provides a reasonable qualitative picture of the electronic structure sufficient for studying band-filling corrections. In this context, the absolute value of the band gap is less important than the relative positions of the energy levels and the behavior of the electrons within them.

**Acknowledgment:** The authors thank the "ENSEMBLE[3] - Centre of Excellence for nanophotonics, advanced materials and novel crystal growth-based technologies" project (GA No. MAB/2020/14) carried out within the International Research Agendas programme of the Foundation for Polish Science co-financed by the European Union under the European Regional Development Fund and the European Union's Horizon 2020 research and innovation programme Teaming for Excellence (GA. No. 857543) for support of this work. We gratefully acknowledge Poland's high-performance computing infrastructure PLGrid (HPC Centers: ACK Cyfronet AGH) for providing computer facilities and support within computational grant no. PLG/2023/016228.

**DATA AVAILABILITY**
The data that support the findings of this study are available from the corresponding author upon reasonable request.